\documentclass[twocolumn,showpacs,preprintnumbers,amsmath,amssymb]{revtex4}
\usepackage{graphicx}    
\usepackage{dcolumn}    
\usepackage{bm}              
\usepackage{amsmath,amsfonts,amssymb}
\usepackage{graphicx,epic,eepic}
\usepackage{subfigure}
\usepackage{graphics}
\usepackage[dvips]{color}
\usepackage{epsfig}
\usepackage{units}      
\usepackage{mathrsfs}  
\usepackage[dvips]{color}
\usepackage{verbatim}

\begin{document}
%%%%%%%%%%%%%%%%%%%%%%%%%%%%%%%%%%%%%%%%%%%%%%%%%%%
% \preprint{gr-qc/xxx}
%%%%%%%%%%%%%%%%%%%%%%%%%%%%%%%%%%%%%%%%%%%%%%%%%%%
%%%%%%%%%%%%%%%%%%%%%%%%%%%%%%%%%%%%%%%%%%%%%%%%%%%
\title{Generalised superradiant scattering}
%%%%%%%%%%%%%%%%%%%%%%%%%%%%%%%%%%%%%%%%%%%%%%%%%%%
%%%%%%%%%%%%%%%%%%%%%%%%%%%%%%%%%%%%%%%%%%%%%%%%%%%
\author{Mauricio Richartz$^*$}
%\email{richartz@phas.ubc.ca}
%
\author{Silke Weinfurtner$^{*\dag}$}
%\email{silke@phas.ubc.ca}
%
\author{A.~J.~Penner$^*$}
%\email{ajpenner@phas.ubc.ca}
%
\author{W.~G.~Unruh$^*$}
%\email{unruh@phas.ubc.ca}
%
\affiliation{$^*$Department of Physics and Astronomy, University of British Columbia,
6224 Agricultural Road, Vancouver, B.C. V6T 1Z1 Canada}
\affiliation{$^\dag$Astrophysics Sector, International School for Advanced Studies, Via Beirut
2-4, 34014 Trieste, Italy and INFN, Sezione di Trieste}
%
%%%%%%%%%%%%%%%%%%%%%%%%%%%%%%%%%%%%%%%%%%%%%%%%%%%%
% \date{13 December 2007; Revised XX; \LaTeX-ed \today}
%%%%%%%%%%%%%%%%%%%%%%%%%%%%%%%%%%%%%%%%%%%%%%%%%%%
\begin{abstract}
We analyse the necessary and sufficient conditions for the occurance of superradiance. Starting with a wave equation we examine the possibility of superradiance in terms of an effective potential and boundary conditions. In particular, we show that the existence of an ergoregion is not sufficient; an appropriate boundary condition, e.g. only ingoing group velocity waves at an event horizon, is also crucial. After applying our scheme to the standard examples of superradiance, we show that analogue models of gravity without an event horizon do not necessarily exhibit superradiance. Particularly, we show that the superradiant phenomenon is absent in purely rotating inviscid fluids with vorticity. We argue that there should be a catalogue of superradiant systems that can be found by focusing on the necessary and sufficient conditions outlined below. 
\end{abstract}
%%%%%%%%%%%%%%%%%%%%%%%%%%%%%%%%%%%%%%%%%%%%%%%%%%%
%%%%%%%%%%%%%%%%%%%%%%%%%%%%%%%%%%%%%%%%%%%%%%%%%%%
\maketitle
%%%%%%%%%%%%%%%%%%%%%%%%%%%%%%%%%%%%%%%%%%%%%%%%%%%
%
%%%%%%%%%%%%%%%%%%%%%%%%%%%%%%%%%%%%%%%%%%%%%%%%%%%
%
%\section{Motivation\label{Sec:Superradiance.Scattering.Nature}}
%
%%%%%%%%%%%%%%%%%%%%%%%%%%%%%%%%%%%%%%%%%%%%%%%%%%%
Wave scattering processes are characterized by the interaction between an incident wave and a physical obstacle, e.g., light rays scattering off raindrops to form a rainbow and x-ray scattering used in medical imaging. 
 In standard scattering processes, incident waves lose energy due to interaction with the media they traverse. Their incoming amplitude is greater than the amplitude of the reflected waves. However, we know of the existence of some special systems where this behaviour is reversed. The amplitude of the reflected wave is larger than the amplitude of the incoming one, meaning energy is extracted from the system. The most popular examples of this phenomenon, known as superradiance~\cite{1988AnPhy.181..261M}, are the scattering of scalar waves by rotating black holes~\cite{1973JETP...37...28S,1973ZhETF..65....3S}, and the scattering of electromagnetic waves by a rotating cylinder made of electrical conductive material~\cite{1972JETP...35.1085Z,Bekenstein:1998nt}.

In this paper, we investigate the details behind the phenomenum of superradiance. In our set-up, the wave scattering process is described by a second order differential equation and an effective potential determined by the interaction between the incident wave and the scattering obstacle. We consider a very general potential and derive the necessary and sufficient condition to invoke superradiant wave scattering. Counterintuitively, such scattering only happens in systems where there is absorption, e.g.~a electrically conductive rotating cylinder or a rotating black hole.
Starting with the standard examples and generalising to other systems, we show that superradiance is possible when the imaginary part of the effective potential is negative. We also demonstrate that the existence of an ergoregion, i.e.~a region where no physical observer can remain at rest, is not sufficient for the ocurrence of superradiance; an appropriate boundary condition is also required. We show the relevance of our result for some acoustic spacetime geometries, and point out some misconceptions appearing in the recent literature where an ergoregion was present but the boundary condition was imposed without clear reasoning~\cite{Lepe:2004kv,Slatyer:2005ty}. Finally, we indicate the difference between superradiance and dynamical instabilities.

%%%%%%%%%%%%%%%%%%%%%%%%%%%%%%%%%%%%%%%%%%%%%%%%%%%
%
%\section{Nature of superradiance\label{Sec:Superradiance.Scattering.Nature}}
%
%%%%%%%%%%%%%%%%%%%%%%%%%%%%%%%%%%%%%%%%%%%%%%%%%%%
Let $(t,\eta,\boldsymbol{\chi})$ be a coordinate system where $t$ is the time coordinate and $(\eta,\boldsymbol{\chi})$ are spatial coordinates. Suppose that a scattering process can be described by a separable field, $\Psi(t,\eta,\boldsymbol{\chi})=h(\eta)\,g(\boldsymbol{\chi})e^{-i\omega t}$, where $\omega \in \mathbb{R}$ and $h(\eta)$ obeys an homogeneous linear second order differential equation. By an appropriate change of variables, this equation can always be cast in the following form,
\begin{equation} \label{waveeq}
\frac{d^2 f}{d \xi ^2} + [V(\xi) + i\,\Gamma(\xi)] f = 0,
\end{equation}
where $f$ is the new dependent variable, $\xi=\xi(\eta)$ is the new independent coordinate and the effective potentials $V(\xi), \Gamma(\xi) \in \mathbb{R}$. We further assume that, when $\xi \rightarrow \infty$,
\begin{equation} 
V(\xi) \rightarrow \omega ^2 \quad \text{and} \quad  \ \xi \, \Gamma(\xi) \rightarrow 0, 
\end{equation}
which admits a solution corresponding to the scattering of an incident wave from infinity,
\begin{equation} \label{fasymp}
f(\xi) = e^{-i \omega \xi} + \mathcal{R}e^{+i \omega \xi}, \quad  \xi \rightarrow \infty,
\end{equation}
 where the amplitude of the reflected wave, $\mathcal{R}$, is called the reflection coefficient. In order to investigate the occurance of superradiance, we need a conservation relation, which can be obtained by considering the spatial derivative of the Wronskian \footnote{The physical interpretation of equation (\ref{wronskian}) depends on the system we are analysing. For example, for a Klein--Gordon field, the Wronskian $i \, W(f,f^*)$ can be interpreted as a particle current in the $\xi$ direction.} of $f(\xi)$ and $f^*(\xi)$,
\begin{eqnarray} 
\frac{d}{d \xi} \left[ i \, W(f,f^*) \right] &= & i\frac{d}{d \xi} \left[ f \frac{d f^*}{d \xi} - f^* \frac{d f}{d \xi}  \right] = 2 \, \Gamma \, |f|^2.
\end{eqnarray}
Integrating the above equation from a point $\xi_0$ (where a boundary condition is imposed) to $\infty$ and plugging in the asymptotic form of $f(\xi)$, see equation~(\ref{fasymp}), we obtain,
\begin{equation} \label{wronskian} 
|\mathcal{R}|^2 = 1 + \frac{i}{2 \omega}W(f,f^*)|_{\xi_0} -\frac{1}{\omega} \int_{\xi _0} ^{\infty} \Gamma(\xi)|f(\xi)|^2 d\xi.
\end{equation}

For the systems described above, the necessary and sufficient condition for superradiance, i.e.~$|\mathcal{R}|>1$, is $iW(f,f^*)|_{\xi_0} - 2\int_{\xi _0} ^{\infty} \Gamma(\xi)|f(\xi)|^2 d\xi > 0$ and so a sufficient condition is $iW(f,f^*)|_{\xi_0} \geq 0$ and $\Gamma(\xi) \leq 0, \, \, \forall \, \xi \in (\xi_0,\infty)$ (at least one of the inequalities must be strict). On the other hand, superradiance is impossible if $iW(f,f^*)|_{\xi_0}  \leq  0$ and $\Gamma(\xi) \geq 0, \, \, \forall \, \xi \in (\xi_0,\infty)$. In these cases it is sufficient to know the behaviour of the solution near the boundary $\xi_0$. However, for the most general case, it is necessary to solve the differential equation (\ref{waveeq}) to determine the ocurrance or absence of superradiance.

We will now discuss in detail different physical systems using the conservation equation (\ref{wronskian}). We start with the standard cases of superradiance, i.e. black hole scattering and Zel'Dovich's cylinder, before going a step further to consider analogue spacetimes.

%%%%%%%%%%%%%%%%%%%%%%%%%%%%%%%%%%%%%%%%%%%%%%%%%%%
%
%\section{Standard examples}
%
%%%%%%%%%%%%%%%%%%%%%%%%%%%%%%%%%%%%%%%%%%%%%%%%%%%
%+++++++++++++++++++++++++++++++++++++++++++++++++++++++++++++++++
%\subsection{Charged scalar waves in a KN black hole} 
%+++++++++++++++++++++++++++++++++++++++++++++++++++++++++++++++++
The scattering process of a scalar wave with electric charge $e$ in a Kerr--Newman background (mass $M$, charge $Q$, specific angular momentum $a$) is described by a separable Klein--Gordon field~\cite{PhysRevD.45.532},
\begin{equation}
\psi(t,r,\theta,\phi)=\frac{f(r)}{\sqrt{r^2 + a^2}}e^{-i \omega t}e^{i m \phi} S_{\ell m}(\theta), 
\end{equation}
where $m$ is the azimuthal number index, $\ell$ is the orbital number index and $S_{\ell m}$ are the spheroidal harmonics. By an appropriate change of coordinates, 
\begin{equation}
\frac{d \xi}{dr}=\frac{r^2 + a^2}{\Delta},
\end{equation}
 where $\Delta = r^2 - 2Mr +Q^2 + a^2$, the Klein--Gordon equation reduces to (\ref{waveeq}) with $\Gamma(\xi)=0$ and
\begin{equation} \label{black hole} 
V =   \left( \omega - \frac{am + eQr}{\tilde{r}^2}  \right)^2 - \lambda^2 \frac{ \Delta}{\tilde{r}^4} +\frac{\Delta}{\tilde{r}^3}\frac{d}{dr}\left(\Delta\frac{d}{dr} \frac{1}{\tilde{r}} \right),
\end{equation}
where $\tilde{r}^2=r^2 + a^2$, $\lambda$ is a separation constant, and $r=r(\xi)$. Furthermore,
\begin{equation}
V \rightarrow
\left\{
\begin{array}{ll}
\omega ^2,  & \xi \rightarrow +\infty (r \rightarrow \infty)
\\
(\omega - m\Omega _h - e\Phi _h)^2 , \ \ \ & \xi \rightarrow -\infty (r \rightarrow r_+)
\end{array}
\right. 
\end{equation}
where $r_+=M+\sqrt{M^2 - Q^2 - a^2}$ is the event horizon, while $\Omega _h = a/(r_+ ^2 + a^2)^2$ and $\Phi_h =Qr_+/(r_+ ^2 + a^2)^2$ are the event horizon's angular velocity and electric potential, respectively.
An incoming wave from $+\infty$, described by equation~(\ref{fasymp}), is scattered by the black hole. As nothing can classically escape from it, only ingoing (i.e. group velocity towards the black hole) solutions are allowed near the horizon,
\begin{equation}
f(\xi)=\mathcal{T}e^{-i(\omega -m\Omega_h -e \Phi _h)\xi}, \ \ \ \ \xi \rightarrow -\infty (r \rightarrow r_+), 
\end{equation}
where $\mathcal{T}$ is the transmission coefficient. From the conservation equation (\ref{wronskian}) we obtain,
\begin{equation} \label{superc}
|\mathcal{R}|^2 = 1 -\frac{\omega -m\Omega_h -e \Phi _h}{\omega}|\mathcal{T}|^2.
\end{equation}
This equation reflects the conservation of particle current~\cite{wald}. If $\omega < m\Omega_h + e \Phi _h$, this current is outgoing at the horizon. Consequently, to obey conservation, the particle current at infinity must also be outgoing and, therefore, superradiance must occur. Another way to understand the physics behind the phenomenon is comparing the energies and direction of propagation of the waves. Far from the black hole, group and phase velocities point in the same direction. However, near the horizon, the phase velocity of the ingoing wave points outwards. Consequently, even though the wave is propagating towards the black hole, rotational energy is being extracted from it, as confirmed by an energy flux calculation at the horizon~\cite{wald}.

We would like to emphasise the importance of the boundary condition to the results. Suppose that instead of having an event horizon, the boundary conditions were different, such that outgoing waves with amplitude $\mathcal{Y}$ were also allowed. Expression (\ref{superc}) would then become
\begin{equation}
|\mathcal{R}|^2 = 1 -\frac{\omega -m\Omega_h -e \Phi _h}{\omega}\left(|\mathcal{T}|^2 - |\mathcal{Y}|^2\right).
\end{equation}
and,  for $|\mathcal{T}| > |\mathcal{Y}|$, the condition on $\omega$ for superradiance would be unchanged~\footnote{Notice that, if $|\mathcal{T}| < |\mathcal{Y}|$, one would obtain $|\mathcal{R}|> 1$ for sufficiently large frequencies $\omega > m\Omega_h + e \Phi _h$. This should not be called superradiance since $|\mathcal{T}| < |\mathcal{Y}|$ corresponds to the scattering of the wave with a source.}.
This could also be applied to a rapidly rotating star with an ergoregion. For simplicity, we assume Kerr--Newman to be the exterior metric and the surface of the star to be perfectly reflecting, $|\mathcal{T}| = |\mathcal{Y}|$. Consequently, $|\mathcal{R}| = 1$, see for example~\cite{Kang:1997uw}. 

%+++++++++++++++++++++++++++++++++++++++++++++++++++++++++++++++++
%\subsection{ZelÕDovich's Rotating Cylinder}
%+++++++++++++++++++++++++++++++++++++++++++++++++++++++++++++++++
A different kind of physical system where superradiance occurs is Zel'Dovich's rotating cylinder~\cite{1972JETP...35.1085Z}. Following the analysis and notation of reference~\cite{Bekenstein:1998nt}, we consider an infinitely long cylinder of radius $R$ rotating in vaccum with constant angular velocity $\Omega$. The cylinder has spatially uniform permittivity $\epsilon(\omega) \in \mathbb{R}$, permeability $\mu(\omega) \in \mathbb{R}$, and electrical conductivity $\sigma \geq 0$. We consider axial electric modes with $k=0$, which are characterized by the following electric,
\begin{equation} 
 \mathbf{E} = \frac{\gamma}{\omega}(\omega - m \Omega) \frac{f(r)}{\sqrt{r}}e^{-i\omega t}e^{i m \phi} \boldsymbol{\hat{z}},
\end{equation}
and magnetic fields,
\begin{equation} 
 \mathbf{B} = \left( \frac{\gamma}{\omega r}(m - \omega\Omega r^2) \boldsymbol{\hat{r}} + \frac{i}{\omega}\boldsymbol{\hat{\phi}} \, \frac{d}{dr}  \right) \frac{f(r)}{\sqrt{r}} e^{im\phi} e^{-i\omega t},
\end{equation}
where $m>0$ is the azimuthal index number and $\gamma$ is the Lorentz factor. The radial function $f(r)$ satisfies equation (\ref{waveeq})
with $r=\xi$ and effective potentials,
\begin{equation} %\nonumber
V=
\left\{
\begin{array}{ll} 
\omega ^2 -\frac{ 4m^2 - 1}{4r^2}, & r > R
\\
\omega ^2 + (1-\epsilon \mu)(\omega -m \Omega)^2 \gamma ^2 -\frac{ 4m^2 - 1}{4r^2}, & r < R
\end{array}
\right. 
\end{equation}
and
\begin{equation} 
\Gamma(r)=
\left\{
\begin{array}{ll}
0,  & r > R
\\
4\pi\gamma\mu\sigma(\omega - m\Omega), \quad & r < R
\end{array}
\right. .
\end{equation}
 In the asymptotic limit, a solution to equation~(\ref{waveeq}) is given by~(\ref{fasymp}). Near $r=0$, the only solution for which the electric and magnetic fields are well behaved is
$f \propto \sqrt{r}r^m$.

The corresponding Wronskian appearing in the conservation equation~(\ref{wronskian}) vanishes,
\begin{equation} 
|\mathcal{R}|^2 = 1 -\frac{1}{\omega} \int_{0} ^{R} 4\pi\gamma\mu\sigma(\omega - m\Omega)\frac{|f(r)|^2}{r} dr.
\end{equation}

If the material is an electrical insulator ($\sigma = 0$), nothing happens to the incident wave; it is reflected back without losing or gaining energy ($|\mathcal{R}|=1$). However, if the cylinder is an electrical conductor ($\sigma \neq 0$), dissipation occurs and $|\mathcal{R}| \neq 1$. The energy loss is proportional to the frequency $\omega - m \Omega$ of the incident wave measured in the co-rotating cylinder frame, see for example~\cite{landau}. For $\omega > m\Omega$ energy is dissipated away in form of heat. Otherwise, for $\omega < m\Omega$, rotational energy is being transferred from the cylinder to the electromagnetic wave and superradiance occurs. This is a general result for axisymmetric macroscopic bodies with constant angular velocity $\Omega$ that can internally dissipate absorbed energies. By the second law of thermodynamics, superradiance arises whenever $\omega - m \Omega < 0$, see~\cite{Bekenstein:1998nt}.  

%%%%%%%%%%%%%%%%%%%%%%%%%%%%%%%%%%%%%%%%%%%%%%%%%%%
%
%\section{Analogue spacetimes}
%
%%%%%%%%%%%%%%%%%%%%%%%%%%%%%%%%%%%%%%%%%%%%%%%%%%%
Equation~(\ref{waveeq}) is also applicable in the context of the analogue spacetime programme~\cite{Barcelo:2005fc}. For acoustic rotating black holes, we show that superradiant scattering processes are possible, see~\cite{Basak:2002aw,Basak:2003uj}. However, if the spacetime does not possess an event horizon, we prove that superradiance is absent for the systems we are investigating.  In particular, we demonstrate that superradiance is not possible in purely rotating inviscid fluids with vorticity.

The equations of motion for small perturbations around a $(2+1)$-dimensional irrotational, inviscid and incompressible water flow can be described by a separable massless Klein--Gordon field in a curved background~\cite{Berti:2004ju,Basak:2003uj},
\begin{equation} 
\psi=
\left\{
\begin{array}{ll}
r^{-\frac{1}{2}+i \frac{mB}{A}}(r^2 - A^2)^{i \frac{\omega A}{2} - i \frac{mB}{2A}} f e^{-i \omega t}e^{i m \phi}, & A\neq 0,
\\
r^{-\frac{1}{2}} f e^{-i \omega t}e^{i m \phi}, & A = 0,
\end{array}
\right. 
\end{equation}
where $m$ is the azimuthal number index, and the background velocity profile is,
 \begin{equation} \label{irrot}
 \mathbf{v}_0=-\frac{A}{r}\boldsymbol{\hat{r}} + \frac{B}{r}\boldsymbol{\hat{\phi}}.
 \end{equation}
By an appropriate change of coordinates, 
\begin{equation}
\frac{d\xi}{dr}=\left(1 - \frac{A^2}{r^2} \right)^{-1},
\end{equation}
the radial Klein--Gordon equation reduces to (\ref{waveeq}), with $\Gamma(\xi)=0$ and
\begin{equation} \label{acblack hole} 
V = \left( \omega - \frac{mB}{r^2} \right)^2 - \left(1 - \frac{A^2}{r^2} \right)\left(\frac{m^2}{r^2} - \frac{1}{4r^2} + \frac{5A^2}{r^4} \right).
\end{equation}
When $A\neq 0$, an event horizon is located at $r=A$ ($\xi=-\infty$). The situation is analogous to the black hole case. Imposing only ingoing waves near the horizon, we obtain the previously derived reflection condition~(\ref{superc}) for $\Phi_h=0$ and $\Omega _h =B/ A^ 2$. In the presence of an event horizon, the exact details of the background velocity at the origin are unimportant as the boundary conditions are imposed at the event horizon.
Therefore, the calculations are insensitive to changes in the velocity profile inside the black hole. 

However, when A = 0, an event horizon no longer exists and a boundary condition must be specified at $r=0$. To circumvent this difficulty, a water tank that extends only
from $r=r_0$ to $r=\infty$ can be considered. Since the field must be continuous at $r_0$, the boundary condition $f(r_0)=0$ has to be imposed. The corresponding solution,
\begin{equation}
f(r)=\alpha (r-r_0) + O(r-r_0)^2, \ \ \ \ \alpha \in \mathbb{C},
\end{equation}
implies a vanishing Wronskian in equation~(\ref{wronskian}) and, consequently, $|\mathcal{R}|=1$. This happens even when an ergoregion is present ($B>r_0$), since the negative energy modes are no longer trapped inside an event horizon~\footnote{The absence of superradiance for rotating acoustic black holes (with $|\mathcal{R}|<1$) has been derived in~\cite{Lepe:2004kv}. However the assumption to requiring only ingoing waves near the ergoregion (which were defined using the phase velocity instead of the group veolicity) is unclear to the present authors.}.

One way to avoid a divergent velocity at the origin is to consider a more realistic velocity profile,
$\mathbf{v}_0=u(r)\boldsymbol{\hat{\phi}}$, where
\begin{equation} \label{newprofile}
u(r)\propto
\left\{
\begin{array}{ll}
r^ \alpha,  \quad  & r\rightarrow 0,
\\
r^{-1}, & r\rightarrow \infty,
\end{array}
\right. 
\end{equation}
with $\alpha \geq 1$. 
Consequently, the fluid flow is no longer
irrotational and the perturbative analysis of the hydrodynamical equations is more sophisticated. It is necessary to introduce additional degrees of freedom in the form of a vector field $\boldsymbol{\zeta}$. Altogether we are dealing with two fields, $\psi$ and $\boldsymbol{\zeta}$, that satisfy the following system of differential equations~\cite{PerezBergliaffa:2001nd}
\begin{eqnarray}
 \frac{d^2 \psi}{d t^2}  = \nabla ^2 \psi + \nabla \cdot \boldsymbol{\zeta}, \label{vort1} \\
\frac{d \boldsymbol{\zeta}}{dt}  = \nabla \psi \times \boldsymbol{\omega}_0 - (\boldsymbol{\zeta} \cdot \nabla )\mathbf{v}_0, \label{vort2}
\end{eqnarray}
where 
\begin{equation}
\boldsymbol{\omega}_0 = \nabla \times \mathbf{v}_0 = \frac{1}{r}\frac{d}{dr}\left[r \, u(r) \right]\boldsymbol{\hat{z}} \equiv \omega_0 (r) \boldsymbol{\hat{z}} 
\end{equation} 
is the background vorticity. 
Performing  the following separation of variables, $\psi(t,r,\phi)=R(r)e^{-i \omega t}e^{i m \phi}$ and $\boldsymbol{\zeta}(t,r,\phi)=\boldsymbol{\varsigma}(r)e^{-i \omega t}e^{i m \phi}$,
equation (\ref{vort2}) becomes
\begin{equation} 
\boldsymbol{\varsigma}(r) =\frac{\omega_0}{K\tilde{\omega}} \left[\frac{\boldsymbol{\hat{r}}}{r}  \left( -m  + 2\frac{u}{ \tilde{\omega}}\frac{d}{dr} \right) + i \boldsymbol{\hat{\phi}} \left( \frac{m \omega_0 }{r \tilde \omega }  - \frac{d}{dr} \right) \right] R,
\end{equation}
where 
\begin{equation} \nonumber
\tilde \omega = \omega - m\frac{u(r)}{r} \quad \text{and} \quad K=1-2\frac{ \omega_0 \, u}{\tilde \omega ^2 \, r}. 
\end{equation}
It is possible to cast equation~(\ref{vort1}) into the form of
(\ref{waveeq}), with $\xi=r$, $\Gamma(r)=0$, $f(r)=R(r)\sqrt{r/K}$, and
\begin{equation} 
V=\frac{\tilde \omega ^2}{\tilde c ^2} - \frac{m^2}{r^2} 
-\frac{1}{2} \frac{d^2}{dr^2} \log \left( \frac{r}{K} \right)-\frac{1}{4}\left[ \frac{d}{dr} \log \left( \frac{r}{K} \right) \right]^2,
\end{equation}
where
\begin{equation}
\frac{1}{\tilde c ^2}=K\left[1 - \frac{m}{r\tilde \omega} \frac{d}{dr}\left( \frac{\omega _0}{K\tilde \omega ^2} \right) \right].
\end{equation}
In the asymptotic limit, $r\rightarrow \infty$, the fluid flow~($\ref{newprofile}$) is irrotational. Therefore, $\boldsymbol{\zeta}=0$ and $f(r)$ given by (\ref{fasymp}) describes an incident wave. Near $r=0$, the only solution to~($\ref{waveeq}$) which admits physically reasonable velocity perturbations ($\delta \mathbf v = \nabla \psi + \boldsymbol{\zeta}$) is $f(r)\propto r^{m+1/2}$, for which the Wronskian $W(f,f^*)|_{0}=0$. Plugging this result into equation~(\ref{wronskian}), we conclude that $|\mathcal{R}|=1$ and superradiance does not occur. It would be interesting to include viscosity in these purely rotating fluid flows. Similarly to Zel'Dovich's cylinder, the dissipative term in (\ref{wronskian}) might then contribute to a superradiant scattering.

A similar situation where an ergoregion might be present without an event horizon is a BEC vortex. In order to determine the ocurrance of superradiance, one needs to know the beviour of the field near the vortex. In \cite{Slatyer:2005ty}, it was argued that superradiance is possible. However the reason for assuming only ingoing waves at the vortex core is not clear to the present authors. It also has to be investigated carefully whether the growing influence of the quantum-pressure (accountable for the non-linear part in the excitation spectrum) near the core interferes with the formation of an ergoregion.

%%%%%%%%%%%%%%%%%%%%%%%%%%%%%%%%%%%%%%%%%%%%%%%%%%%
%
%\section{Rotating stars and Instabilities}
%
%%%%%%%%%%%%%%%%%%%%%%%%%%%%%%%%%%%%%%%%%%%%%%%%%%%
Even though superradiance is absent in non-dissipative systems without an event horizon, dynamical instabilities might be present. Those correspond to an unbounded exponential growth in time ($\mathfrak{Im}(\omega)>0$) and the validity of the perturbative treatment eventually breaks down. The ansatz (3) is not appropriate in this case. An additional boundary condition of purely outgoing waves at $\xi=+\infty$ must be imposed and one has to solve an eigenvalue problem for complex frequencies.

Initially stimulated by the so called acoustic spacetime programme, where it is possible to mimic some but not all features of a rotating black hole, we were able to extract the necessary and sufficient conditions for superradiant wave scattering.
The strength of our formalism is twofold: the universality and the applicability. The universality lies in the model-independent description of wave scattering in terms of an effective potential and boundary conditions. One can check, in a straightforward manner, if the necessary and sufficient conditions for superradiance, i.e.~negative imaginary part of the Wronskian and\,/\,or negative dissipation, are fulfilled. Within our framework, it is not necessary to fully solve the original wave equation to obtain a qualitative statement for superradiant scattering. The Wronskian only has to be calculated at the boundary and the sign of the dissipation term can be read off directly.
We have demonstrated this for various known, e.g.~Kerr--Newman black holes and Zel'Dovich's rotating cylinder, and new systems, e.g.~rotating fluids with vorticity. 
In conclusion, superradiant scattering of waves is not an isolated effect that only appears in very few, hard-to-access systems. There are strong indications for the existence of a catalogue of systems that exhibit superradiant scattering processes. 
%%%

% Acknowledgements 

MR was supported by FAPESP and CAPES. SW was supported by a Marie Curie Fellowship EMERGENT-2007-SW. We wish to thank Matt Visser for his comments.

%%%%%%%%%%%%%%%%%%%%%%%%%%%%%%%%%%%%%%%%%%%%%%%%%%%
%%%%%%%%%%%%%%%%%%%%%%%%%%%%%%%%%%%%%%%%%%%%%%%%%%%
\bibliographystyle{apsrev}

%%%%%%%%%%%%%%%%%%%%%%%%%%%%%%%%%%%%%%%%%%%%%%%%%%%
%%%%%%%%%%%%%%%%%%%%%%%%%%%%%%%%%%%%%%%%%%%%%%%%%%%
\end{document}